\documentclass[10pt,conference]{IEEEtran}
\IEEEoverridecommandlockouts
\usepackage[letterpaper, top=0.76in, bottom=1.1in, left =0.7in, right=0.7in]{geometry}
\usepackage{cite}
\usepackage{afterpage}
\usepackage{amsmath,amssymb,amsfonts}
\usepackage{algorithmic}
\usepackage{braket}
\usepackage{graphicx}
\usepackage{textcomp}
\usepackage{xcolor}
\usepackage{tikz, float}
\usepackage{subcaption}   
\usepackage{pgfplots, stmaryrd}
\pgfplotsset{compat=1.18}
\newcommand{\bsc}[1]{BSC\left( #1 \right)}
\def\BibTeX{{\rm B\kern-.05em{\sc i\kern-.025em b}\kern-.08em 
    T\kern-.1667em\lower.7ex\hbox{E}\kern-.125emX}}
\begin{document}

\title{Interpolation of Quantum Polar Codes and Quantum Reed-Muller Codes }

\author{\IEEEauthorblockN{Keita Hidaka}\IEEEauthorblockA{\textit{Faculty of Engineering,}\\\textit{The University of Tokyo,} \\Japan \\keita-hidaka717@g.ecc.u-tokyo.ac.jp}
\and \IEEEauthorblockN{Dina Abdelhadi}\IEEEauthorblockA{\textit{School of Computer } \\ \textit{and Communication Sciences, EPFL,}\\Switzerland\\dina.abdelhadi@epfl.ch}\and\IEEEauthorblockN{Ruediger Urbanke}\IEEEauthorblockA{\textit{School of Computer } \\\textit{and Communication Sciences, EPFL,}\\Switzerland \\rudiger.urbanke@epfl.ch}}

\maketitle

\begin{abstract}
Good quantum error-correcting codes that fulfill practical considerations, such as simple encoding circuits and efficient decoders, are essential for functional quantum information processing systems. Quantum polar codes satisfy some of these requirements but lack certain critical features, thereby hindering their widespread use. Existing constructions either require entanglement assistance to produce valid quantum codes, suffer from poor finite-size performance, or fail to tailor polar codes to the underlying channel properties. Meanwhile, quantum Reed-Muller (RM) codes demonstrate strong performance, though no known efficient decoding algorithm exists for them. In this work, we propose strategies to interpolate between quantum polar codes and quantum RM codes, thus addressing the challenges of designing valid quantum polar codes without entanglement assistance and improving finite-size code performance.
\end{abstract}

\begin{IEEEkeywords}
QEC, Reed-Muller Codes, Polar Codes
\end{IEEEkeywords}

\section{Introduction}
Quantum error correction is a crucial piece of the puzzle of quantum computation and communication. Many quantum error correction approaches rely on building on existing classical error correcting codes.

Of classical codes, polar codes occupy a prominent position. They boast simple encoding circuits and efficient decoders such as the successive cancellation (SC) decoder, and were among the first examples of code constructions achieving the capacity of binary-input memoryless symmetric channels (BMS) \cite{arikan2009channel}. A channel-independent construction using polarization-weight polar codes \cite{zhou2018polarizationweightfamilymethods} has recently been adopted into the 5G standard.

However, it has been noted that the finite-size performance of polar codes suffers from some drawbacks, as large blocklengths are required to achieve reliable communication with a fixed rate and error probability \cite{Korada2010Empirical}. Reed-Muller (RM) codes are a related family of codes, which have also been shown to achieve the capacity of BMS channels \cite{reeves2021reed,abbe2023proof}. Although no efficient decoder is known yet for RM codes, they have been numerically observed to have improved error rates compared to polar codes under MAP decoding \cite{Hussami2009Performance}. It is this observation that inspired the work of \cite{mondelli2014polarreedmullercodestechnique} where families of codes interpolating between RM and polar codes were introduced. By applying successive cancellation list (SCL) decoding to these families of interpolating codes, the authors of \cite{mondelli2014polarreedmullercodestechnique} showed a construction that improves finite-size performance at the cost of a controlled increase in decoding complexity.

Given the impressive success of classical polar codes, it is natural to consider quantum versions of polar codes for the transmission of quantum information. Quantum CSS polar codes have been shown in \cite{Renes2012Efficient} to achieve the symmetric coherent information rate of qubit Pauli and erasure channels. This construction, however, requires pre-shared entanglement, which remains a significant challenge for practical implementations.
In contrast, a different approach in \cite{renes2015efficient} achieves the coherent information without the need for pre-shared entanglement, but employs an involved multilevel encoding scheme, with unclear finite-size performance and a potential need for large blocklengths to ensure reliable communication.
Existing quantum devices are far from the regime where asymptotic behaviour is relevant, with only a few qubits, making good finite-size performance essential for any practical applications. More recently, the polarization-weight method for polar coding has been extended to the quantum domain \cite{gong2024improved}, exhibiting good performance at small to moderate blocklengths, with a simple channel-independent construction.  Moreover, this method is guaranteed to produce a valid quantum code with a commuting set of stabilizers. 

In this work, we propose code constructions interpolating between quantum polar codes (QPC) and quantum Reed-Muller codes (QRM), inspired by the work of \cite{mondelli2014polarreedmullercodestechnique}. Taking the quantum channel parameters into consideration in code design, we show improved performance over the polarization-weight method of \cite{gong2024improved} under SCL decoding with the decoder implementation of \cite{gongPWQPC}.
By varying an interpolating parameter $\alpha$, we scan over the quantum code family indexed by $\alpha$, to identify valid codes that require no pre-shared entanglement, and then select $\alpha$ values that yield outperforming constructions under SCL decoding at practically relevant blocklengths. Our proposed approach offers a tunable tradeoff between decoding complexity and desirable code properties. For example, a large group of automorphisms is desirable, as they have been shown to be very useful for fault-tolerant implementations of quantum gates \cite{Grassl_2013}. While RM codes have a large automorphism group, the general affine group, polar codes possess less symmetry. Yet, \cite[Figure 2]{geiselhart2021automorphismgrouppolarcodes} shows that as polar codes designed for the binary erasure channel (BEC) become more ``RM-like" at lower noise rates, the size of the automorphism group grows. 

\section{Preliminaries}

\subsection{Pauli Channels}\label{sec:pauli}
Pauli noise is a model often used for noise affecting quantum communication and computation. 
Let $$X=\begin{bmatrix}0&1\\1&0\end{bmatrix},Z=\begin{bmatrix}1&0\\0&-1\end{bmatrix}, Y=iXZ$$ denote the Pauli matrices.
Note that $X$, $Z$ anticommute, i.e., $XZ=-ZX$. 
Consider a quantum state of a qubit described by a positive-semidefinite density matrix $\rho \in \mathbb{C}^{2\times 2}, \text{Tr}[\rho]=1$. 
The action of a general Pauli noise channel on $\rho$ is described by the mapping:
 $$\rho \rightarrow (1-p_X-p_Y-p_Z) \rho+p_X X\rho X+p_Y Y \rho Y + p_Z Z\rho Z,$$ where $p_X,p_Y,p_Z \geq 0,$ and $ p_X+p_Y+p_Z \leq 1.$
In this work, we focus on the noise model of independent, equal $XZ$ noise given by: 
   $p_X =p_Z= q-q^2,\quad p_Y= q^2,$ for some $0\leq q\leq 1.$
Recalling \cite[Section 1.5.2]{goswami2021quantum}, we associate with each Pauli channel two classical binary input memoryless (BMS) channels: 

\begin{enumerate}
    \item Bit-flip channel $W_X:\bsc{p_X + p_Y}$,
    \item Phase-flip channel $W_Z$: a probabilistic mixture of BSCs along with an output flag indicating if an $X$-error has occurred,
    \[
    \left\{
    \begin{aligned}
        &\bsc{\frac{p_Y}{p_X + p_Y}}, && \textnormal{flag} = 1, && \textnormal{(prob.\ } p_X + p_Y\textnormal{)} \\
        &\bsc{\frac{p_Z}{p_I + p_Z}}, && \textnormal{flag} = 0, && \textnormal{(prob.\ } p_I + p_Z\textnormal{)}
    \end{aligned}
    \right.
    \]
    This mixture expresses the phase-flip channel observed given the outcome of decoding the error on the bit-flip channel and the correlation between $X$ and $Z$-type errors.
\end{enumerate}

The ability to decode and correctly estimate the errors on these two induced classical channels implies the ability to correct the Pauli errors on the quantum Pauli channel. 

\subsection{Quantum CSS codes}
A quantum error correcting code designed for a Pauli channel needs to correct two types of errors: $X$ and $Z$ errors. Since $Y=iXZ$, $Y$ errors are considered to be the simultaneous occurrence of $X$ and $Z$ errors. In a classical linear block code, syndrome information enabling the detection and correction of errors is obtained by applying or measuring parity checks. Measuring parity checks corresponds to measuring \emph{stabilizers} of a stabilizer quantum code. 

The main challenge in the quantum case is that stabilizer measurements correspond to observables that must commute to be measured simultaneously. 
Quantum CSS codes are a large subclass of quantum stabilizer codes, which provide a straightforward method for constructing quantum error correcting codes based on classical codes, guaranteeing the construction of a valid commuting set of stabilizers.  
For CSS codes, the stabilizer generators fall into two categories: $X$-type $\in \{I,X\}^{\otimes N}$ and $Z$-type $\in \{I,Z\}^{\otimes N}$. Stabilizers of the same type always commute. Determining if an $X$-type stabilizer commutes with a $Z$-type stabilizer can be done by taking the dot product (over $GF(2)$ of the corresponding binary strings, constructed as $I\rightarrow 0, Z\rightarrow 1, X \rightarrow 1$. If the dot product is 0, the two strings have an even number of occurrences of $X$ and $Z$ at corresponding positions, so the stabilizers commute; otherwise, they anticommute.

A quantum CSS code is constructed from two linear binary classical codes $C_1=[N,k_1],C_2=[N,k_2]$, such that $C_1^\perp \subseteq C_2$. The rows of the parity check matrix of $C_1$, $H_1$, is used to construct $Z$-type stabilizer generators, with the mapping $0\rightarrow I, 1\rightarrow Z$. Similarly, the parity check matrix $H_2$ of $C_2$ is used to construct $X$-type stabilizer generators. Since $C_1^\perp \subseteq C_2$ implying $H_1 H_2^T = 0$ (over $GF(2)$), the constructed stabilizers are guaranteed to commute, yielding a quantum code with parameters $\llbracket N, k_1+k_2-N\rrbracket$.

\section{Polar \& RM Codes}
Let $E = \begin{bmatrix}
    1&0\\
    1&1
\end{bmatrix}$. For a code of blocklength $N=2^n$, denote the polar transform as $G=E^{\otimes n}$. Note that the polar transform satisfies $GG=I$ over $GF(2)$.
The generators of the codewords of classical polar and RM codes are chosen as subsets of rows of $G.$ The distinguishing feature between polar and RM codes is how these subsets are selected.

An RM code is defined via two integer parameters: $m$, where the code blocklength is $N=2^m$, and $r$, where the code minimum distance is $2^{m-r}$. To construct RM codes of minimum distance $2^{m-r}$, the codeword generators are selected as the rows of $G$ of Hamming weight $\geq 2^{m-r}$. As a result of this construction, the family of RM codes has a nested structure, where $RM(m,r)\subset RM(m,r^\prime)$ whenever $r^\prime > r$. Note that the RM code construction does not depend on channel parameters. It is also useful to note that the dual of an $RM(m,r)$ code is another Reed-Muller code $RM(m,m-r-1).$

When constructing polar codes, a virtual channel is defined for each input index. Then, a parameter evaluating the quality of each virtual channel is computed. The rows of $G$ which generate the codewords are those rows with indices corresponding to the best virtual channels according to the computed channel parameter. 

Thus, for both polar and RM codes, the code is defined via a set of information indices $\mathcal{I}$, and the complementary set of indices is called the frozen set $\mathcal{F}$.

\subsection{Constructing Polar Codes}
The polar transform $G$ is applied to an input vector $u$, yielding $x = uG$, where each bit of $x$ is transmitted over $N$ i.i.d. instances of a noisy channel $W$, yielding a noisy output vector $y$ at the receiver.
The virtual channels are defined via operations called channel splitting and channel combining, the details of which can be found in \cite{arikan2009channel,goswami2021quantum}. 
There are $N$ such channels, where the $i^{th}$ channel is a map from a binary input bit $u_i$ to an output vector of length $N+i$ consisting of the output $y$, as well as (the estimates of) the previous $u_0^{i-1}$ bits. This virtual channel is denoted by $W^{(i)}(y,u_0^{i-1}|u_i)$. 
There are many possible metrics for evaluating the quality of the virtual channel, for example: 
\begin{itemize}
    \item Bhattacharyya parameter $Z(W^{(i)})$, given by $$\sum_{y,u_0^{i-1}\in \{0,1\}^{N+i}} \sqrt{W^{(i)}(y,u_0^{i-1}|0)W^{(i)}(y,u_0^{i-1}|1)} ,$$ 
    \item $P(\mathcal{E}_i)$ is the probability that the $i^{th}$ bit is decoded incorrectly by the SC decoder, where the decoder decides the bit $u_i$ according to the sign of the log-likelihood ratio: $l_i(y,u_0^{i-1}) = \log \frac{W^{(i)}(y,u_0^{i-1}|0)}{W^{(i)}(y,u_0^{i-1}|1)}$. Thus, 
    $$P(\mathcal{E}_i) = \text{Pr}\left[U_i \neq \frac{1-\text{sign}(l_i(Y,U_0^{i-1}))}{2}\right].$$
    In \cite{arikan2009channel}, it is shown that $P(\mathcal{E}_i)$ is upper-bounded by the Bhattacharyya parameter of $W^{(i)}$ and that the overall probability of SC decoder error is upper-bounded by $\sum_i P(\mathcal{E}_i)$.
\end{itemize}
Note that the alphabet size of the output of the virtual channels $W^{(i)}(y,u_0^{i-1}|u_i)$ grows exponentially with block size, making the exact evaluation of the channel quality parameters inefficient. To remedy this issue, \cite{tal2013construct,pedarsani2011construction} provide an approximation algorithm that computes tight upper and lower bounds on the channel parameters.

\subsection{Interpolations of Codes}
Order the virtual channels of a polar code assuming that the underlying physical channel is BEC$(\alpha\epsilon)$. Construct a code of rate $R$ by picking the fraction $R$ of best channels according to this ordering. Here $R$ is of the order of $1-\epsilon$ depending on code length to take into account the finite-length effects. Clearly, if $\alpha=1$ we just get the standard polar code for the BEC$(\epsilon)$. Perhaps slightly more interestingly, it was shown in \cite{mondelli2014polarreedmullercodestechnique} that if we take the limit $\alpha \rightarrow 0$ then we get the RM code of the same rate. 

This observation led to the definition of a family of interpolating codes $\{\mathcal{C}_\alpha\}_{\alpha \in [0,1]}$ as the polar codes designed for BEC$(\alpha\epsilon)$ at a fixed rate, where SCL decoding of these codes showed promising results at practical list sizes \cite{mondelli2014polarreedmullercodestechnique}. In the same work, extensions of these interpolations are also discussed for BMS channels. There are many ways to define the interpolating code families, leading to many variants of such families. 

\section{Quantum Polar Codes and Quantum RM Codes}
Quantum polar codes (QPC) and quantum RM codes (QRM) can both be constructed from their classical counterparts
by representing the polar transform $G$ as a circuit with controlled-not (CNOT) gates, and replacing classical CNOT gates with quantum CNOT gates to turn it into a quantum circuit.

To convert the quantum operation into two classical operations, we focus on the effect of quantum CNOT on $X$ and $Z$ errors. 
A quantum CNOT gate maps $I\otimes X \rightarrow
I\otimes X$, $X\otimes I \rightarrow X\otimes X$, similar to the action of a classical CNOT gate under the mapping $(I\leftrightarrow 0, X\leftrightarrow 1)$. On the other hand, it maps $I\otimes Z \rightarrow Z\otimes Z$, $Z\otimes I \rightarrow
Z\otimes I$, which indicates that quantum CNOT acts on Pauli strings $\in \{Z,I\}^{\otimes n}$ in the reverse order compared to its action on Pauli strings $\in \{X,I\}^{\otimes n}$. As a result of the quantum polar transform, the virtual channels synthesized from the bit-flip channel $W_X$ are polarized according to the polarization matrix $G$, while the virtual channels synthesized from the phase-flip channel $W_Z$ are polarized according to the polarization matrix with the input and output orders reversed, given by $RGR = G^T$, where $R$ is a matrix that applies the reversal. 

For each qubit input index $i$, two channel parameters are computed. The first parameter for the bit-flip channel is equal to the parameter of the $i^{th}$ virtual channel synthesized from the classical channel $W_X$ according to the classical polar transform $G$. The second parameter for the phase-flip channel is equal to the parameter of the $(N-i-1)^{th}$ virtual channel synthesized from the classical channel $W_Z$ according to the classical polar transform $G$ (or equivalently that of the $i^{th}$ channel synthesized according to $G^T$).

For QPC, the channel parameters can be one of the channel metrics for classical polar codes, such as Bhattacharyya parameter or $P(\mathcal{E}_i)$. 
To fit the QRM code construction into the same framework, we take the two channel parameters for the $i^{th}$ input qubit for QRM codes to be the Hamming weights of the $i^{th}$ and the $(N-i-1)^{th}$ row of $G$, where a higher Hamming weight indicates a better channel.

To construct a quantum code with a certain fixed rate $R$, we decide on the classical code rates $k_1, k_2$ such that $NR = k_1+k_2-N$. Then, the best $k_1$ channels according to the channel parameters of $W_X$ are selected, while the rest of the channel indices form the frozen set $\mathcal{F}_Z$,  with their inputs fixed in the $Z$-basis to $\ket{0}$. The best $k_2$ indices according to the computed channel parameters of $W_Z$ are selected, while the remaining channel indices form the frozen set $\mathcal{F}_X$,  with their inputs fixed in the $X$-basis to $\frac{\ket{0}+\ket{1}}{2}$.

Once the set of input indices frozen in the $Z$-basis $\mathcal{F}_Z$ and the set of input indices frozen in the $X$-basis $\mathcal{F}_X$ are determined, one can construct the stabilizers of the code as follows: 
\begin{itemize}
    \item $Z$-type stabilizers ($\in \{I,Z\}^{\otimes N}$) are the columns of the matrix $G$ indexed by the set $\mathcal{F}_Z$,
    \item $X$-type stabilizers ($\in \{I,X\}^{\otimes N}$) are the columns of the matrix $G^T$ indexed by the set $\mathcal{F}_X$ (equivalently, the rows of the matrix $G$ indexed by the set $\mathcal{F}_X$).
\end{itemize}
 To check the conditions for these stabilizers to commute, consider an index $a \in \mathcal{F}_Z$ and an index $b \in \mathcal{F}_X$. Computing the dot product of the $a^{th}$ row and $b^{th}$ column of $G$ to check the commutation, we get $\sum_{i,j} G_{a,j}G_{i,b}$. Due to the fact that  $GG=I$, $\sum_{i,j} G_{a,j}G_{i,b}=\delta_{a,b}$. Thus, to ensure that the stabilizers commute, no index should be in both $ \mathcal{F}_Z$ and $ \mathcal{F}_X$. 
 
 This can be guaranteed for a QRM code with a positive rate, where ``good" channels are those with Hamming weight of at least some threshold value.
 Let the blocklength $N=2^m$. The Hamming weight of the $i^{th}$ row of $G$ has been shown \cite[Proposition 17]{arikan2009channel} to be $=2^{\sum_{j} b_j^{(i)}}$, where $ b_j^{(i)}$ is the $j^{th}$ bit of the binary representation of $i$.
If $i \in \mathcal{F}_X$, then it must be that the Hamming weight of the $i^{th}$ row is below some threshold integer value $2^w$, i.e., ${\sum_{j} b_j^{(i)}} \leq w-1$. 
Similarly, if $i \in \mathcal{F}_Z$, then it must be that ${\sum_{j} b_j^{(N-i-1)}} = {m-\sum_{j} b_j^{(i)}}< {w'}$ . Assume for the sake of contradiction that $i \in \mathcal{F}_X \cap \mathcal{F}_Z$, then $m < w+w'-1.$ Since the QRM code is assumed to have a positive rate, 

\begin{align*}
    &k_1+k_2-N > 0 \\
    \Leftrightarrow &\sum_{j=0}^{m-w} \binom{m}{j}-\sum_{j=0}^{ w'-1}\binom{m}{j}>0,
\end{align*}
which implies $m-w>w'-1$ . Thus, we get a contradiction, and conclude that $\nexists i \in  \mathcal{F}_X \cap \mathcal{F}_Z$. 
This proof applies to QRM codes constructed such that channels corresponding to indices of a certain Hamming weight are either frozen or not, i.e., there cannot be two indices $i,j$ such that both $i,j$ have equal Hamming weight but $i$ is frozen and $j$ is not. 
This enforces some constraint on allowed code dimensions, where $k_1 = \sum_{l=0}^{r_1} \binom{m}{l}, k_2= \sum_{l=0}^{r_2} \binom{m}{l}$, for some integers $r_1,r_2$. We obtain more flexible code parameters using the RM construction of \cite{gong2024improved}, where the index $i$ is assigned an RM virtual channel parameter given by $\sum_j b_j^{(i)}+i/N$, breaking ties between rows of equal Hamming weights.
The statement that $\nexists i \in  \mathcal{F}_X \cap \mathcal{F}_Z$ is not guaranteed for QPC. 
\section{$\alpha$-QPC-QRM interpolated codes}
For our proposed code constructions, we rely on identifying the classical BMS channels corresponding to the Pauli channel, as described in Section \ref{sec:pauli}. 
We can then create the interpolating code families by first defining the corresponding interpolating classical BMS channels: 

\begin{enumerate}
    \item $W_{X}^\alpha:\bsc{\alpha(p_X+p_Y)}$,
    \item $W_Z^\alpha$: a probabilistic mixture of BSCs along with an output flag,\[
    \left\{
    \begin{aligned}
        &\bsc{\alpha\frac{p_Y}{p_X + p_Y}}, && \textnormal{flag} = 1, && \textnormal{(prob.\ } p_X + p_Y\textnormal{)} \\
        &\bsc{\alpha\frac{p_Z}{p_I + p_Z}}, && \textnormal{flag} = 0, && \textnormal{(prob.\ } p_I + p_Z\textnormal{)}
    \end{aligned}
    \right.
    \]
\end{enumerate}
In the case of independent, equal $XZ$ noise, these channels reduce to two copies of $\bsc{\alpha q}$, where $p_X =p_Z = q-q^2, p_Y =q^2$.
We fix $k_1, k_2$ for the target quantum CSS code that we wish to construct for a Pauli channel with noise parameter $q$, and compute the virtual channel parameters for $\bsc{\alpha q}$ with several values of $\alpha \in [0, 1]$. In our implementation, $P(\mathcal{E}_i)$ is estimated using the approximation algorithm discussed in \cite{tal2013construct,pedarsani2011construction}.
Then, the indices are sorted according to the computed channel parameters. The indices corresponding to the worst $N-k_1$ channels are frozen in the $Z$-basis to create the set $\mathcal{F}_Z(\alpha)$. 
For the bits frozen in the $X$-basis, the list of channel parameters is reversed, such that the channel parameter of the $i^{th}$ virtual phase-flip channel is equal to the 
channel parameter of the $(N-i-1)^{th}$ virtual bit-flip channel. 
The indices corresponding to the worst $N-k_2$ channels are frozen in the $X$-basis to create the set $\mathcal{F}_X(\alpha)$. 
We then check which values of $\alpha$ correspond to sets $\mathcal{F}_Z(\alpha)$, $\mathcal{F}_X(\alpha)$ such that $\nexists i \in  \mathcal{F}_X \cap \mathcal{F}_Z$. Those values produce valid QPC constructions, and are passed on to the SCL decoder. We search over $\alpha$ values and find the ones with the lowest logical error rates under SCL decoding.

\begin{figure}[t]
    \centering
    \begin{tikzpicture}
\definecolor{darkgray176}{RGB}{176,176,176}
\definecolor{darkorange25512714}{RGB}{255,127,14}
\definecolor{steelblue31119180}{RGB}{31,119,180}
    \begin{axis}[
tick align=outside,
tick pos=left,
x grid style={darkgray176},
xlabel={$q$},
xmajorgrids,
xtick style={color=black},
ymode=log,
 xtick={0.03, 0.04, 0.05, 0.06, 0.07,0.08,0.09,0.1}, 
xticklabel style={/pgf/number format/fixed},
y grid style={darkgray176},
ylabel={Logical error rate under \texttt{SCL-C}},
ymajorgrids,
ytick style={color=black},
legend style={
    at={(0.95,0.1)}, 
    anchor=south east,
    draw=none,
    fill=white,
    draw = black,
    font=\small,
},
point meta=explicit symbolic,
nodes near coords,
nodes near coords align={horizontal},
every node near coord/.append style={font=\tiny, anchor=west},
]        
\addplot[very thick,densely dashed, red, mark=* ] table [x=p,y=alpha,meta=label,]{all_plot.txt};
\addplot[thick,mark options={solid},mark=x, densely dotted, gray ] table [x=p,y=alpha1]{alpha1.txt};
\addplot[very thick, steelblue31119180, mark=o] table [x=p,y=PW]{all_plot.txt};
\addlegendentry{$\alpha$-QPC-QRM};
\addlegendentry{QPC};
\addlegendentry{PW-QPC}
\end{axis}
\end{tikzpicture}
    \caption{The logical $X$ error rate of $\llbracket 1024,42 \rrbracket$ codes with different constructions. Each data point is evaluated by averaging over $10^6$ samples with an \texttt{SCL-C} decoder of list size $16$. For each $q$, we search over multiple $\alpha$ values and show the best performing codes on the plot. The $\alpha$ values of the corresponding codes are in Table \ref{tab:alpha_star}.}
    \label{fig:alpha_result}
\end{figure}
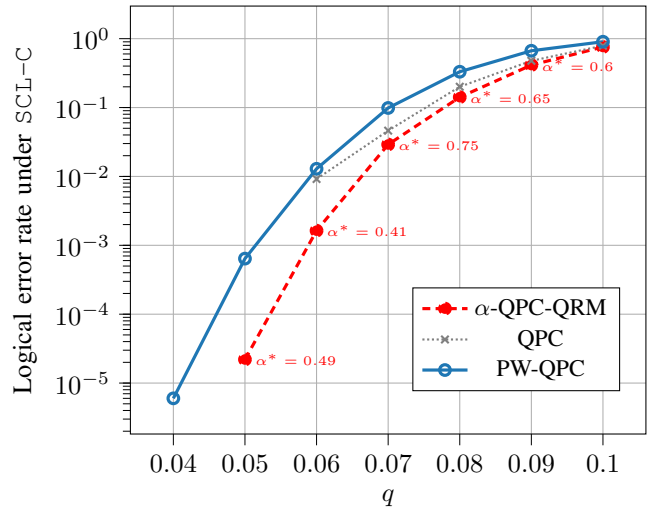
\begin{figure}[!b]
    \centering
    \begin{tikzpicture}
\definecolor{darkgray}{RGB}{176,176,176}
\definecolor{darkorange}{RGB}{255,127,14}
\definecolor{steelblue}{RGB}{31,119,180}
    \begin{axis}[
tick align=outside,
tick pos=left,
x grid style={darkgray},
xlabel={$\alpha$},
xmajorgrids,
ymode=log,
xtick style={color=black},
xticklabel style={/pgf/number format/fixed},
y grid style={darkgray},
ylabel={Logical error rate under \texttt{SCL-C}},
ymajorgrids,
ytick style={color=black},
legend style={at={(0.95,0.95)}, anchor=north east},
]        
\addplot[black, mark=*] table [x=alpha,y=l1]{alpha-L.txt};
\addplot[ thick,darkgray,  mark=o ] table [x=alpha,y=l4]{alpha-L.txt};
\addplot[ thick,darkorange, mark=+ ] table [x=alpha,y=l8]{alpha-L.txt};
\addplot[ thick,steelblue, mark=x ] table [x=alpha,y=l16]{alpha-L.txt};
      \addlegendentry{$L=1$};
      \addlegendentry{$L=4$};
      \addlegendentry{$L=8$};
      \addlegendentry{$L=16$};
    \end{axis}
    \end{tikzpicture}
    \caption{Logical $X$ error rate is simulated at $q=0.0425$, for $\llbracket 1024,94 \rrbracket$}
    \label{fig:list_size_effect}
\end{figure}
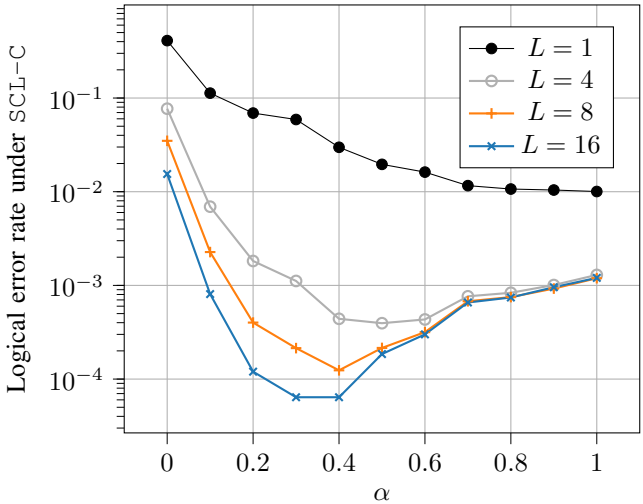

\begin{table*}[t]
    \captionsetup{width=0.9\textwidth}
    \centering\resizebox{0.9\textwidth}{!}{
    \begin{tabular}{c||c|c|c|c|c|c|c}
   & $q=0.04$ & $q=0.05$ & $q=0.06$ & $q=0.07$ & $q=0.08$ & $q=0.09$ & $q=0.10$ \\
    \hline
     \small{$P_{e,\texttt{SCL-C}}(\alpha = 1)$} 
     & invalid & invalid 
    & 0.009176 & 0.046212 & 0.200462 & 0.477210 & 0.791207 \\

   \small{$P_{e,\texttt{SCL-C}}(\alpha^*)$ }
    & $\sim 0$ & $2.2 \times 10^{-5}$ 
    & 0.001632 & 0.029084 & 0.14154 & 0.413467 & 0.757872\\
     $\alpha^*$ 
    & 0.61 & 0.49 & 0.41 & 0.75 & 0.65 & 0.6 & 0.6 \\
    \end{tabular}}
    \caption{Logical error rates at $\alpha=1$ and at best performing values $\alpha^*$, for different values of $q$, averaged over $10^6$ samples  . The results are reported for codes with $N=1024, k_1=k_2 =533, L=16$ decoded using the \texttt{SCL-C} decoder. Whenever $\alpha=1$ does not correspond to a valid quantum code, we denote that as ``invalid". }
    \label{tab:alpha_star}
\end{table*}
\begin{table*}[t]
    \captionsetup{width=0.9\textwidth}
    \centering\resizebox{0.9\textwidth}{!}{
    \begin{tabular}{c||c|c|c|c|c|c|c}
   & $q=0.04$ & $q=0.05$ & $q=0.06$ & $q=0.07$ & $q=0.08$ & $q=0.09$ & $q=0.10$ \\\hline
    $\alpha^*$ 
    & 0.61 & 0.49 & 0.41 & 0.75 & 0.65 & 0.6 & 0.6 \\
     \small{mixing factor} 
     & 414 & 414
    & 414 & 406 & 406 & 406 &406
    \end{tabular}}
    \caption{Mixing factors at best performing values $\alpha^*$, for different values of $q$. The results are reported for codes with $N=1024, k_1=k_2 =533$. }
    \label{tab:mixing_factor}
\end{table*}
\paragraph{Results}
To evaluate the performance of the construction, we use the successive cancellation list decoder (\texttt{SCL-C}) of \cite{gong2024improved,gongPWQPC}. Note that for the independent $XZ$ channel with equal $X,Z$ noise, the logical error rates are identical for the $X$ (bit-flip) and $Z$ (phase-flip) channels. Thus, the results in the plots of Figures \ref{fig:alpha_result}, \ref{fig:list_size_effect} are simulated for $X$-type errors only, representing the logical $X$ error rate.

For each noise parameter $q$, we pick 10 random values of $\alpha$ that produce a valid quantum code, and evaluate the performance of the $\texttt{SCL-C}$ decoder to decide the best value of $\alpha^*$. 
Figure \ref{fig:alpha_result} shows a comparison of the proposed $\alpha$-QPC-QRM codes and PW-QPC \cite{gong2024improved} codes under \texttt{SCL-C} decoding with list size $L=16$, for the independent, equal $XZ$ Pauli noise model . We choose to compare against the best performing code of \cite[Figure 5]{gong2024improved} at  blocklength $N=1024$, $k_1 = k_2 = 533$, with polarization weight parameter $\beta = 2^{1/4}-0.12$. At these parameters, the code has dimension $k=42$. It is observed that a significant improvement in code performance can be achieved by suitably tuning the $\alpha$ parameter.
\paragraph{Finite-size performance gains over regular polar codes}
Note that even when $\alpha=1$ results in a valid quantum polar code, it is usually beneficial to construct an interpolating code with a smaller $\alpha$, to improve the finite-size performance of the code. This can be observed in Table \ref{tab:alpha_star}.

\paragraph{Mixing factors}
The mixing factor is the number of information indices that are smaller than the index of the last frozen bit. It has been shown in \cite{hashemi2017partitionedlistdecodingpolar,fazeli2021list} to be related to the performance of SCL decoding, where a smaller mixing factor indicates that a smaller list size suffices to achieve ML decoding performance. Thus, it is a useful metric of code quality. For the PW-QPC construction $\llbracket 1024, 42\rrbracket$ shown in Fig. \ref{fig:alpha_result}, the mixing factor is 470. The mixing factors of the various $\alpha^*$ codes shown in Figure \ref{fig:alpha_result} are listed in Table \ref{tab:mixing_factor}
\paragraph{Effect of list-size} 
Here we focus on the performance of $\alpha$-QPC-QRM codes of blocklength $N=1024$, $k_1=k_2=559$ yielding a quantum code of higher dimension $k=94$. At this high rate, many values of $\alpha$ lead to valid quantum code constructions. For $q=0.0425$, we vary $\alpha$ and the list size $L$, and observe the value of the resulting \texttt{SCL-C} decoder logical error rate, as shown in Figure \ref{fig:list_size_effect}. We note that the $\alpha$ value corresponding to the best performance decreases as $L$ increases. This is reasonably expected, as $L\rightarrow \infty$ corresponds to MAP decoding, and $\alpha\rightarrow 0$ corresponds to RM codes. It is known that RM codes outperform polar codes under MAP decoding \cite{mondelli2014polarreedmullercodestechnique}.
\paragraph{Evolution of size of the automorphism group with $\alpha$}
As discussed in \cite{geiselhart2021automorphismgrouppolarcodes,DBLP:journals/corr/abs-2103-14215}, the automorphism group of Reed-Muller codes of blocklength $2^n$ is the general affine group $GA(2,n)$, while for polar codes that are decreasing monomial codes, the automorphism group is called the Block Lower-Triangular Affine Group (BLTA).  A large group of automorphisms is desirable, as code automorphisms have
been shown to be  useful for fault-tolerant implementations of quantum
gates \cite{Grassl_2013}. We use the algorithm provided in \cite{geiselhart2021automorphismgrouppolarcodes} (assuming the constructed codes are decreasing monomial codes) to compute how the size of the automorphism group varies with changing $\alpha$. Focusing on the bit-flip errors over an independent $XZ$ channel with $q = p_X+p_Y = 0.06,$ we investigate the automorphism group of the classical polar codes designed for $\bsc{\alpha q}$, at values where $\alpha$ leads to a valid quantum code with parameters $\llbracket 1024, 252 \rrbracket$. We find that the size of the automorphism group for the RM code is $\sim 10^{15} - 10^{16}$ times larger than that of the interpolating codes. However, a modest increase in the size of the automorphism group from $3.6028797\times 10^{16}$ at $\alpha=1$ to $1.0808639\times 10^{17}$ at $\alpha= 0.1$ is observed. 
\section*{Acknowledgements}
The authors would like to express their gratitude to Professor Emre Telatar for explaining and providing the code for approximating the channel parameters. The authors would also like to thank Anqi Gong for interesting discussions and helpful clarifications.

\bibliographystyle{IEEEtran}  
\bibliography{refs}

\begin{thebibliography}{10}
\providecommand{\url}[1]{#1}
\csname url@samestyle\endcsname
\providecommand{\newblock}{\relax}
\providecommand{\bibinfo}[2]{#2}
\providecommand{\BIBentrySTDinterwordspacing}{\spaceskip=0pt\relax}
\providecommand{\BIBentryALTinterwordstretchfactor}{4}
\providecommand{\BIBentryALTinterwordspacing}{\spaceskip=\fontdimen2\font plus
\BIBentryALTinterwordstretchfactor\fontdimen3\font minus
  \fontdimen4\font\relax}
\providecommand{\BIBforeignlanguage}[2]{{%
\expandafter\ifx\csname l@#1\endcsname\relax
\typeout{** WARNING: IEEEtran.bst: No hyphenation pattern has been}%
\typeout{** loaded for the language `#1'. Using the pattern for}%
\typeout{** the default language instead.}%
\else
\language=\csname l@#1\endcsname
\fi
#2}}
\providecommand{\BIBdecl}{\relax}
\BIBdecl

\bibitem{arikan2009channel}
E.~Arikan, ``Channel polarization: A method for constructing capacity-achieving
  codes for symmetric binary-input memoryless channels,'' \emph{IEEE
  Transactions on information Theory}, vol.~55, no.~7, pp. 3051--3073, 2009.

\bibitem{zhou2018polarizationweightfamilymethods}
\BIBentryALTinterwordspacing
Y.~Zhou, R.~Li, H.~Zhang, H.~Luo, and J.~Wang, ``Polarization weight family
  methods for polar code construction,'' 2018. [Online]. Available:
  \url{https://arxiv.org/abs/1805.02813}
\BIBentrySTDinterwordspacing

\bibitem{Korada2010Empirical}
S.~B. Korada, A.~Montanari, E.~Telatar, and R.~Urbanke, ``An empirical scaling
  law for polar codes,'' in \emph{2010 IEEE International Symposium on
  Information Theory}, 2010, pp. 884--888.

\bibitem{reeves2021reed}
G.~Reeves and H.~D. Pfister, ``Reed-{M}uller codes achieve capacity on bms
  channels,'' \emph{arXiv preprint arXiv:2110.14631}, vol.~4, no.~6, p.~7,
  2021.

\bibitem{abbe2023proof}
E.~Abbe and C.~Sandon, ``A proof that {R}eed-{M}uller codes achieve shannon
  capacity on symmetric channels,'' in \emph{2023 IEEE 64th Annual Symposium on
  Foundations of Computer Science (FOCS)}, 2023, pp. 177--193.

\bibitem{Hussami2009Performance}
N.~Hussami, S.~B. Korada, and R.~Urbanke, ``Performance of polar codes for
  channel and source coding,'' in \emph{2009 IEEE International Symposium on
  Information Theory}, 2009, pp. 1488--1492.

\bibitem{mondelli2014polarreedmullercodestechnique}
\BIBentryALTinterwordspacing
M.~Mondelli, S.~H. Hassani, and R.~Urbanke, ``From polar to reed-muller codes:
  a technique to improve the finite-length performance,'' 2014. [Online].
  Available: \url{https://arxiv.org/abs/1401.3127}
\BIBentrySTDinterwordspacing

\bibitem{Renes2012Efficient}
\BIBentryALTinterwordspacing
J.~M. Renes, F.~Dupuis, and R.~Renner, ``Efficient polar coding of quantum
  information,'' \emph{Phys. Rev. Lett.}, vol. 109, p. 050504, Aug 2012.
  [Online]. Available:
  \url{https://link.aps.org/doi/10.1103/PhysRevLett.109.050504}
\BIBentrySTDinterwordspacing

\bibitem{renes2015efficient}
J.~M. Renes, D.~Sutter, F.~Dupuis, and R.~Renner, ``Efficient quantum polar
  codes requiring no preshared entanglement,'' \emph{IEEE Transactions on
  Information Theory}, vol.~61, no.~11, pp. 6395--6414, 2015.

\bibitem{gong2024improved}
A.~Gong and J.~M. Renes, ``Improved logical error rate via list decoding of
  quantum polar codes,'' in \emph{2024 IEEE International Symposium on
  Information Theory (ISIT)}.\hskip 1em plus 0.5em minus 0.4em\relax IEEE,
  2024, pp. 2496--2501.

\bibitem{gongPWQPC}
\BIBentryALTinterwordspacing
A.~Gong, ``Pw-qpc-list-decoder: List decoder for the polarization weight family
  of quantum polar code, github,'' 2023. [Online]. Available:
  \url{https://github.com/gongaa/PW-QPC}
\BIBentrySTDinterwordspacing

\bibitem{Grassl_2013}
\BIBentryALTinterwordspacing
M.~Grassl and M.~Roetteler, ``Leveraging automorphisms of quantum codes for
  fault-tolerant quantum computation,'' in \emph{2013 IEEE International
  Symposium on Information Theory}.\hskip 1em plus 0.5em minus 0.4em\relax
  IEEE, Jul. 2013, p. 534–538. [Online]. Available:
  \url{http://dx.doi.org/10.1109/ISIT.2013.6620283}
\BIBentrySTDinterwordspacing

\bibitem{geiselhart2021automorphismgrouppolarcodes}
\BIBentryALTinterwordspacing
M.~Geiselhart, A.~Elkelesh, M.~Ebada, S.~Cammerer, and S.~ten Brink, ``On the
  automorphism group of polar codes,'' 2021. [Online]. Available:
  \url{https://arxiv.org/abs/2101.09679}
\BIBentrySTDinterwordspacing

\bibitem{goswami2021quantum}
A.~Goswami, ``Quantum polar codes,'' Ph.D. dissertation, Universit{\'e}
  Grenoble Alpes, 2021.

\bibitem{tal2013construct}
I.~Tal and A.~Vardy, ``How to construct polar codes,'' \emph{IEEE Transactions
  on Information Theory}, vol.~59, no.~10, pp. 6562--6582, 2013.

\bibitem{pedarsani2011construction}
R.~Pedarsani, S.~H. Hassani, I.~Tal, and E.~Telatar, ``On the construction of
  polar codes,'' in \emph{2011 IEEE international symposium on information
  theory proceedings}.\hskip 1em plus 0.5em minus 0.4em\relax IEEE, 2011, pp.
  11--15.

\bibitem{hashemi2017partitionedlistdecodingpolar}
\BIBentryALTinterwordspacing
S.~A. Hashemi, M.~Mondelli, S.~H. Hassani, R.~Urbanke, and W.~J. Gross,
  ``Partitioned list decoding of polar codes: Analysis and improvement of
  finite length performance,'' 2017. [Online]. Available:
  \url{https://arxiv.org/abs/1705.05497}
\BIBentrySTDinterwordspacing

\bibitem{fazeli2021list}
A.~Fazeli, A.~Vardy, and H.~Yao, ``List decoding of polar codes: How large
  should the list be to achieve ml decoding?'' in \emph{2021 IEEE International
  Symposium on Information Theory (ISIT)}.\hskip 1em plus 0.5em minus
  0.4em\relax IEEE, 2021, pp. 1594--1599.

\bibitem{DBLP:journals/corr/abs-2103-14215}
\BIBentryALTinterwordspacing
Y.~Li, H.~Zhang, R.~Li, J.~Wang, W.~Tong, G.~Yan, and Z.~Ma, ``The complete
  affine automorphism group of polar codes,'' \emph{CoRR}, vol. abs/2103.14215,
  2021. [Online]. Available: \url{https://arxiv.org/abs/2103.14215}
\BIBentrySTDinterwordspacing

\end{thebibliography}
\appendix
The way our proposed $\alpha$ codes interpolate can be observed in Figure \ref{fig:frozenset_polar} and Figure \ref{fig:frozenset_rm}. 
Let $f_{polar}$, $f_{RM}$ denote the proportion of frozen indices of an $\alpha$ code that overlaps with those of polar code and RM code, respectively, defined as follows:
$$f_{polar}(\alpha) = \frac{|{F}_Z(\alpha) \cap \mathcal{F}_Z(\text{1})|}{| \mathcal{F}_Z(\text{1})|}$$
$$f_{RM}(\alpha) =\frac{|\mathcal{F}_Z(\alpha) \cap \mathcal{F}_Z(\text{0})|}{|\mathcal{F}_Z(\text{0})|}$$
Recall that when $\alpha=0$, the interpolating code is an RM code, and when $\alpha =1$, it is a polar code. These two metrics indicate how much "polar-like" and "RM-like" the produced code is, and help us visualize the interpolation. In Figure \ref{fig:frozenset_polar} and \ref{fig:frozenset_rm}, $f_{polar}$ and $f_{RM}$ for $q=0.03$, $\llbracket 1024,252 \rrbracket$ are plotted as a function of $\alpha$. 

Note that care must be taken to define this overlap properly if the RM code does not include all rows of Hamming weight higher than a fixed value $w$, i.e., if there is some form of tie-breaking between two indices corresponding to rows of equal Hamming weight, where one index is frozen and the other is not. The RM code $\llbracket 1024,252 \rrbracket$ discussed in the plot of Figure \ref{fig:frozenset_polar} and \ref{fig:frozenset_rm} is constructed from two classical RM codes $[1024,638] = RM(m=10,r=5)$ such that all indices of rows of the generating matrix $G$ of Hamming weight $\geq 2^{5}$ are not in the frozen set.
\begin{figure}[t]
    \centering

    \begin{tikzpicture}
\definecolor{darkgray}{RGB}{176,176,176}
\definecolor{darkorange}{RGB}{255,127,14}
\definecolor{steelblue}{RGB}{31,119,180}
    \begin{axis}[width=1.0\linewidth,
tick align=outside,
tick pos=left,
x grid style={darkgray},
xlabel={$\alpha$},
xmajorgrids,
xtick style={color=black},
xticklabel style={/pgf/number format/fixed},
y grid style={darkgray},
ylabel={{$f_{polar}$}},
ymajorgrids,
ytick style={color=black},
legend style={at={(1.05,1)}, anchor=north west},
]        
\addplot[red, mark=+] table [x=alpha,y=polar_like]{interpolation.txt};

    \end{axis}\end{tikzpicture}
    \caption{The proportion of frozen indices of $\alpha$ codes shared with polar code.}
    \label{fig:frozenset_polar}
\end{figure}
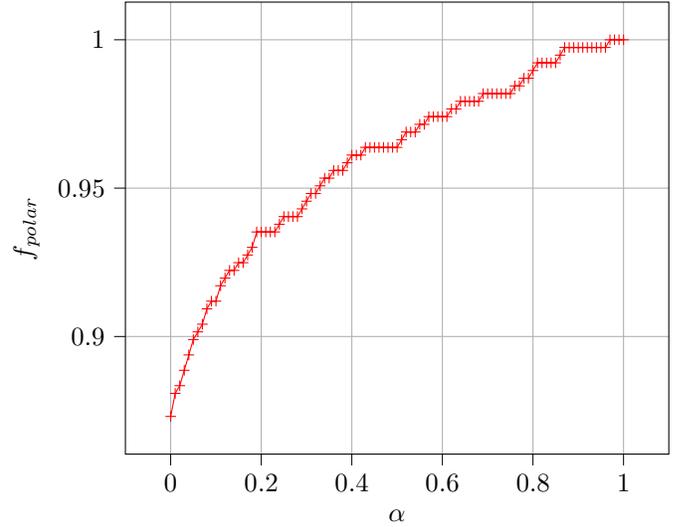
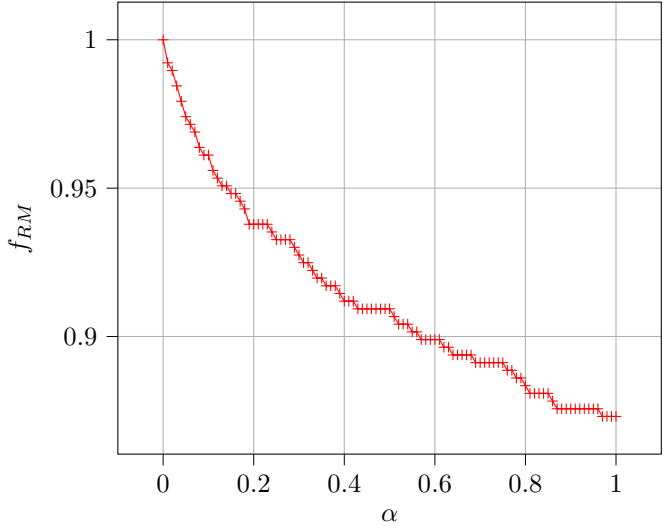
\begin{figure}[t]
    \centering
    \begin{tikzpicture}
\definecolor{darkgray}{RGB}{176,176,176}
\definecolor{darkorange}{RGB}{255,127,14}
\definecolor{steelblue}{RGB}{31,119,180}
    \begin{axis}[width=1.0\linewidth,
tick align=outside,
tick pos=left,
x grid style={darkgray},
xlabel={$\alpha$},
xmajorgrids,
xtick style={color=black},
xticklabel style={/pgf/number format/fixed},
y grid style={darkgray},
ylabel={{$f_{RM}$}},
ymajorgrids,
ytick style={color=black},
legend style={at={(1.05,1)}, anchor=north west},
]        
\addplot[red, mark=+] table [x=alpha,y=RM_like]{interpolation.txt};
    \end{axis}\end{tikzpicture}
    \caption{The proportion of frozen indices of $\alpha$ codes shared with RM code.}
    \label{fig:frozenset_rm}
\end{figure}

The evolution of $f_{polar}$ with decreasing $\alpha$ in Figure \ref{fig:frozenset_polar} can explain the reason why the performance of the $L = 1$ decoder (SC decoder) deteriorates monotonically as $\alpha$ decreases in Figure \ref{fig:list_size_effect}. When we construct an $\alpha$-QPC-QRM interpolated code with $\alpha=1$, its channel parameters are polar code parameters, which gives us the upper bound of the SC decoding error. In other words, the upper bound of the SC decoding error (or its approximation) is minimized under $\alpha=1$. According to Figure \ref{fig:frozenset_polar}, as $\alpha$ decreases, more low-quality virtual channels are included in the information set $\mathcal{I}$, while the corresponding high-quality virtual channels are excluded, thereby degrading the SC decoding performance. This observation suggests that improved performance cannot be expected from the proposed interpolation framework under SC decoder alone, and that combining the SC decoder with list-decoding is crucial. 
Nevertheless, the interpolating codes under SC decoding may still offer advantages, particularly in cases where the QPC construction yields anticommuting stabilizers, by enabling the construction of codes that do not require entanglement assistance.

\end{document}